\documentclass[pdflatex,sn-mathphys-num]{sn-jnl}

\usepackage{graphicx}%
\usepackage{multirow}%
\usepackage{amsmath,amssymb,amsfonts}%
\usepackage{amsthm}%
\usepackage{mathrsfs}%
\usepackage[title]{appendix}%
\usepackage{xcolor}%
\usepackage{textcomp}%
\usepackage{manyfoot}%
\usepackage{booktabs}%
\usepackage{algorithm}%
\usepackage{algorithmicx}%
\usepackage{algpseudocode}%
\usepackage{listings}%
\usepackage{caption} 
\usepackage{graphicx}
\usepackage{subcaption} 
\usepackage{subcaption}
\usepackage{tabularx}
\usepackage{url}
\usepackage{hyperref}  
\usepackage{natbib}
\usepackage{amssymb}
\usepackage{amsmath}
\usepackage{amsthm}
\usepackage{algorithm}
\usepackage{algpseudocode}
\usepackage{booktabs}
\usepackage{siunitx}
\usepackage{subcaption}
\usepackage{caption}
\usepackage{graphicx}
\usepackage{xr}
\usepackage{float}
\usepackage{hyperref}
\usepackage{cleveref}

\theoremstyle{thmstyleone}%
\theoremstyle{thmstyletwo}%

\theoremstyle{thmstylethree}%

\begin{document}

\title[Article Title]{Convolutional Neural Network Optimization for
Beehive Classification Using Bioacoustic Signals}


\author*[1]{\fnm{Harshit} \sur{}}\email{harshitwork4032@gmail.com}
\author[2, 3]{\fnm{Rahul} \sur{Jana}}\email{i.rahuljana@gmail.com}
\author*[2]{\fnm{Ritesh} \sur{Kumar}}\email{riteshkr@csio.res.in}



\affil[1]{
\orgname{National Institute of Technology},
\orgaddress{\street{Thanesar},
\city{Kurukshetra},
\postcode{136119},
\state{Haryana},
\country{India}}}

\affil[2]{
\orgname{CSIR-Central Scientific Instruments Organization},
\orgaddress{
\city{Chandigarh},
\postcode{160030},
\country{India}}}

\affil[3]{
\orgdiv{TIF - Agriculture and Water Technology Development Hub(AWaDH)},
\orgname{Indian Institute of Technology},
\orgaddress{\city{Bara Phool},
\postcode{140001},
\state{Punjab},
\country{India}}}



\abstract{The behavior of honeybees is an important ecological phenomenon not only in terms of honey and beeswax production but also due to  the proliferation of flora and fauna around it. The best way to study this significant phenomenon is by non-invasive monitoring of beehives using the sounds produced by various body movements that give  out audio signals which can be exploited for various predictions related to the objectives mentioned above. This study investigates the application of Convolutional Neural Networks to classify and monitor  different hive states with the help of joint time and frequency image representations such as Spectrogram, Mel-Spectrogram, Smoothed-Spectrogram, and Cochleagram. Our findings indicate that the Cochleagram outperformed all the other representations, achieving an accuracy of 98.31\% on unseen data. Furthermore, we employed various strategies including pruning, quantization, and knowledge distillation to optimize the network and prevent any potential issues with model size. With these optimizations, the network size was lowered by 91.8\% and the inference time was accelerated by 66\%, increasing its suitability for real-time applications. Thus our study emphasizes the significance of using optimization approaches to minimize model size, avoid deployment problems, and expedite inference for real-time application as well as the selection of an appropriate time-frequency representation for optimal performance.}


\maketitle

\section{Introduction}\label{sec1}

Bees are essential to our ecosystem, playing a crucial role in pollination~\cite{beePollination2021} and honey production. In an effort to enhance honey production and ensure the sustainability of the population, artificial  or man-made beehives have become increasingly important. For better production and regulating the bee population, queen bee plays a crucial role~\cite{shaara2021honeybee}. The queen bee’s ability to lay eggs and regulate the hive's social structure directly impacts the health and productivity of the colony~\cite{winston1991biology}. Proper monitoring of these dynamics is crucial for maintaining a healthy hive, especially in artificial environments. Following the importance of the queen bee in maintaining the productive hive we explored audio signals which can be used to classify and monitor the behavior of the hive by analyzing the sound produced by the bees within the hive~\cite{cejrowski2018queen}. Studies have shown that different hive states produce distinct acoustic signatures, which can be effectively classified using advanced machine learning techniques such as Convolutional Neural Networks (CNNs)~\cite{abdollahi2022acoustics}. This approach not only helps in early detection of hive issues but also aids in optimizing hive management practices~\cite{zacepins2016swarming}.

Convolutional neural networks have been showing remarkable results in understanding image classification problems~\cite{liu2019semantic} and now being widely applied to other domains, including signal classification~\cite{Cetin2021}. Since CNNs utilize images of fixed dimensions as input, it has become really essential to wisely choose the techniques to represent the audio signal into an image. For this, we experimented four different time–frequency representation coupled with the CNN including Spectrogram, Mel–Spectrogram, Smoothed–Spectrogram and Cochleagram. As shown in the paper by Haran at el., cochleagram time–frequency image representation fits the best with CNN~\cite{haran2021benchmarking}, and we witnessed the same that the cochleagram time–frequency image representation outperformed all the other representations. And after hypertuning, network improvement and data augmentation by transforming the raw audio files ( e.g. pitch shift, time stretch, and change speed)~\cite{paiva2022} we achieved an accuracy of 98.31\% on unseen data.

Although the performance of the Cochlea–CNN is outstanding, but the size and inference time can cause deployment issues, exploitation of resources and may cause lag in the real–time operation because of larger size and slow inference respectively as compared to traditional statistical machine learning approach. In our research we explored different size reduction and inference acceleration techniques including pruning, knowledge distillation and quantization and implemented them subsequently while preserving the performance of the Cholea–CNN. After optimization, we significantly reduced the size of the model by 91.8\% from 21.82 mb to 1.79 mb and accelerated the inference by 66\%. Our research shows the role of choosing the appropriate time-frequency representation to achieve high performance and the potential of optimization techniques for reducing the size of the model to avoid any deployment issues, resources exploitation and to accelerate inference to avoid lag in the real–time process.

The remainder of this paper is organized as follows. Section~\ref{sec:related_work} lists a review of the related work in bioacoustic signal classification and beehive monitoring and classification. Section ~\ref{sec:methodology} details the proposed methodology for benchmarking and optimization, including data preprocessing, feature extraction, classification approach, and model optimization techniques for reducing neural network size and inference time. Sections ~\ref{sec:results} \&~\ref{sec:discussion}  shows the experimental  results, and an in-depth performance analysis and discussion. Finally, Section ~\ref{sec:conclusion} concludes the paper and outlines potential directions for future research.

\section{Related Work}\label{sec:related_work}
Several studies have explored sound-based monitoring on beehives. For instance, a proposed method for identifying queen bees using audio analysis and machine learning was developed~\cite{Nolasco2019}. In their study, audio data was sourced from a public dataset provided by the Nu-Hive project~\cite{Labonno2022}. In their work, SVM and CNN were the experimented models and SVM performed better than CNN with an average test AUC score of approximately 0.94. The author performed experiments using SVMs, incorporating various feature sets such as 120 frequency bands from 85 Mel spectra, 20 MFCCs, and 20 bands from the Hilbert–Huang Transform (HHT). They also exploited MFCCs and Mel-spectrograms coupled with CNN for feature extraction, and the resulting features were then input into the SVM classifier, achieving an AUC of approximately 0.82.

Another study explored the effectiveness of various algorithms in machine learning, including Support Vector Machines (SVM), Random Forest (RF), and Extreme Gradient Boosting (XGBoost), as well as two CNN architectures—shallow CNN and VGG-13—for feature extraction in identifying bee and non-bee sounds~\cite{cejrowski2018queen}. In their study, they used audio data from beehives provided by the Open Source Beehives (OSBH) initiative, which is included in a publicly available dataset~\cite{Nolasco2019}. Their findings showed that VGG-13, coupled with Mel Frequency Cepstral Coefficients (MFCCs) as input, achieved an accuracy of 91.93\%. Moreover, VGG-13 proved most effective at identifying non-bee sounds, with an F1-score of 0.79.

Similarly, the research utilized CNN coupled with MFCC wave spectrogram images for identifying the beehive sound~\cite{winston1991biology}; this approach achieved 85.04\% accuracy.

Soares et al.~\cite{Soares} explored an MFCC–based descriptor using machine learning and deep learning methods to identify queen bee presence by analyzing the spectral and temporal characteristics of hive acoustics. The author utilized publicly available datasets related to queen bee presence, specifically incorporating audio recordings from the Open Source Beehive project and the NU-Hive dataset~\cite{Labonno2022}, with annotations provided by Nolasco and Benetos~\cite{Nolasco2018}. A total of 508 samples, each averaging 10 minutes in duration, were processed into 15-second segments. These segments were converted into Mel-spectrograms on a logarithmic scale for CNN input, using a sampling rate of 22,050 Hz, window size of 4096, hop length of 2030, and the Hann window function~\cite{Labonno2022}.

They evaluated four CNN architectures that were pre-trained on the ImageNet dataset~\cite{Russakovsky2015} to extract deep features: ResNet50~\cite{He2015}, VGG16~\cite{zhang2015}, VGG19~\cite{zhang2015}, and Inception\_v3~\cite{Szegedy2016}. In their work, handcrafted features were chosen to distinguish between the presence and absence of the queen bee, and recursive feature elimination with cross-validation (RFECV)~\cite{Koul2019} was applied to refine the feature set. The algorithm identified 36 significant features, which included 31 MFCC coefficients, two contrast measures, zero-crossing rate (ZCR), polynomial coefficients, and spectral flatness. For classification, the author explored the Random Forest (RF) and Support Vector Machine (SVM) classifiers and fed the extracted features to them. The best classification performance using deep features was achieved with the ResNet50 model combined with Random Forest (RF), reaching an accuracy of 0.99, while the combination of MFCC features with RF also yielded an accuracy of 0.99. However, their work only focuses on binary classification — either the queen bee is present or not.

However, to our knowledge, no one has yet worked or experimented with cochleagram time-frequency representation for beehive classification and multiclass beehive state classification. In our research, we explored the potential of cochleagram time–frequency representation, outperforming all other time–frequency representations including spectrogram, mel–spectrogram, and smoothed–spectrogram, showing the potential of CNNs in classifying audio signals. In addition, the dataset was divided into four classes: Queen Not Present, Queen Present and Newly Accepted, Queen Present and Rejected, Queen Present or Original Queen.

\section{Methodology}\label{sec:methodology}
This section outlines the detailed methodology implemented to develop and optimize a CNN for classifying beehive states from bioacoustic data is shown. We started with the conversion of raw bioacoustic signals into spectrograms, Mel-spectrograms, smoothed-spectrograms, and cochleagrams time–frequency representations. These representations are used as fixed dimension input to Convolutional Neural Networks were employed to extract features from the time-frequency representation of bioacoustic data. We then benchmarked the effectiveness of each image representation was assessed to determine the most suitable one with the CNNs. Subsequently, we describe the classification model and evaluation metrics which are used to validate and evaluate the model. For Further enhancing the efficiency, we applied three model optimization techniques, pruning, knowledge distillation, and quantization. These strategies significantly reduce model/network size and inference time while preserving the model’s accuracy and effectiveness. The following subsections detail each step, from feature extraction to network compression.
\subsection{Data Collection }

The bees audio data we have used was collected from kaggle.com~\cite{anna_yang_2022}. The dataset, is the largest of its kind which consists of beehive audio data paired with multi-dimensional sensor data~\cite{abdollahi2022acoustics}. It was gathered utilizing a specially designed IoT device equipped with an ESP32 Wi-Fi module, an INMP441 microphone, and a BME280 sensor to measure temperature and humidity. The recordings were obtained from European honeybee colonies located in California and is divided into 60-second segments, resulting in a total of 7,100 samples. It was divided into four classes Queen Not Present, Queen Present and Newly Accepted, Queen Present and Rejected, Queen Present or Original Queen~\cite{winston1991biology}.

\subsection{Data Preprocessing}
\begin{figure}[h]
    \centering
    \begin{subfigure}[t]{0.48\textwidth}
        \centering
        \fbox{\includegraphics[width=0.95\linewidth, height=3cm]{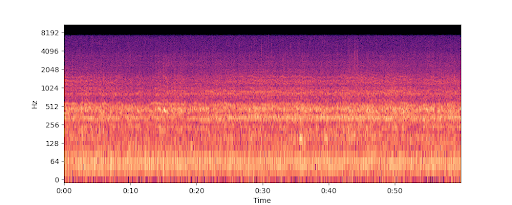}}
        \caption{} \label{fig:spectrogram}
    \end{subfigure}
    \hfill
    \begin{subfigure}[t]{0.48\textwidth}
        \centering
        \fbox{\includegraphics[width=0.95\linewidth, height=3cm]{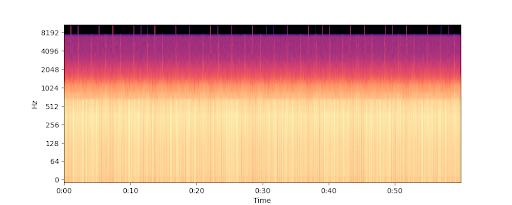}}
        \caption{} \label{fig:smoothed}
    \end{subfigure}
    \hfill
    \begin{subfigure}[t]{0.48\textwidth}
        \centering
        \fbox{\includegraphics[width=0.95\linewidth, height=3cm]{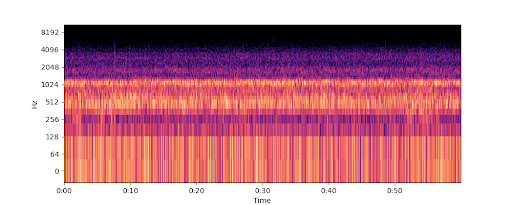}} 
        \caption{} \label{fig:mel} 
    \end{subfigure}
    \hfill
    \begin{subfigure}[t]{0.48\textwidth}
        \centering
        \fbox{\includegraphics[width=0.95\linewidth, height=3cm]{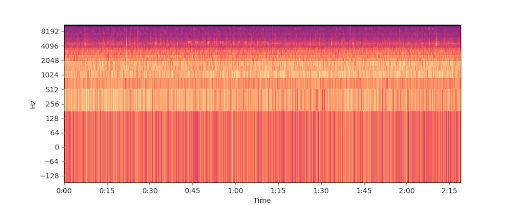}}
        \caption{} \label{fig:cochleagram}
    \end{subfigure}

    \caption{Comparison of time–frequency image representations of a beehive bioacoustic signal using different transformation techniques. (a) Spectrogram, (b) Smoothed Spectrogram, (c) Mel–Spectrogram, and (d) Cochleagram. These representations are used as input to the CNN model for multiclass hive state classification.}
    \label{fig:time_series_rep}
\end{figure}

It has become really important to process raw data since we were using CNN for classification, and it consumes images of fixed input to perform further computations~\cite{Ghosh2019}. In our research, we experimented with four different time-frequency image representations of raw audio data files including spectrogram, mel-spectrogram, smoothed-spectrogram, and cochleagram as shown in ``Fig.~\ref{fig:spectrogram},~\ref{fig:smoothed},~\ref{fig:mel} and ~\ref{fig:cochleagram}'' respectively. We adopted LibROSA library~\cite{McFee2015} to generate spectrogram, Mel–Spectrogram, and Smoothed–spectrogram and the Cochleagram representations were generated using gtgram module present in the gammatone library~\cite{Heeris2013}

\subsubsection{Spectrogram}\label{aa}
Spectrograms have been produced through the application of the Short-Time Fourier Transform (STFT)~\cite{Oppenheim1999}. It first breaks the audio signals in overlapping frames, and then DFT was applied to each frame to convert them to frequency domain~\cite{Oppenheim1999,Smith2007}. Formula for segmenting the signal into overlapping frames and DFT for each frame is shown in equation~\ref{eq:1}, where x[n] represents the discrete time sample of the signal, n is the length of each frame,  m is the step size between frames ( usually m \(<\) n) and w[n] is a window function applied on each frame to minimize spectral leakage.~\cite{Oppenheim1999,Muller2015}

\begin{equation}
x_m[n]=x[n+mM]\cdot w[n]
\label{eq:1}
\end{equation}

DFT can be computed using the equation~\ref{eq:2}.

\begin{equation}
X(k, r) = \sum_{n=0}^{N-1} x(n) w(n) e^{-2\pi i \frac{k n}{N}}, \quad k = 0, \dots, N-1 \label{eq:2}
\end{equation}

Where, \textit{X(k,r)} is the k-th harmonic for the r-th frame~\cite{Harris1978}.
The spectrogram values are then obtained by taking the logarithm of the magnitude of the DFT values:

\begin{equation}
 S(k, r) = \log \left| X(k, r) \right|^2 \label{eq:3}
\end{equation}

The number of points for FFT and the hope length of the STFT was 1024 and 512 respectively. These parameters determine the time-frequency resolution of the spectrogram and are commonly used in signal processing tasks ~\cite{Oppenheim2010}.

\subsubsection{Mel–Spectrogram}
he Mel-spectrogram represents a modified form of a spectrogram where the frequency axis is transformed according to the Mel scale. It was obtained after computing STFT using equation~\ref{eq:3}, then the magnitude spectrogram was converted to a power spectrogram using equation~\ref{eq:4} before passing it through the Mel–Filterbank~\cite{fayek2016,McFee2015}. 

\begin{equation}
P(k, r) = \left| S(k, r) \right|^2 \label{eq:4}
\end{equation}

For a given power spectrogram P(m,k), the Mel-Spectrogram M(m,n) is obtained using the equation~\ref{eq:5}.

\begin{equation}
M(k, r) = \sum_{n=0}^{N-1} H_r(n) \cdot P(k, n) \label{eq:5}
\end{equation}

where $H_r(n)$ signifies the weight of the r-th Mel filter for the n-th frequency bin, and N denotes the total count of frequency bins in the spectrogram~\cite{OShaughnessy1987}.  The count of points for FFT and the hope length was the same as in the spectrogram.

\subsubsection{Smoothed–Spectrogram}
It follows the same process that was used to generate spectrograms with additional smoothing by taking the moving average along the time and frequency axis~\cite{Patterson1992}. Equation~\ref{eq:6},~\ref{eq:7} is used for time and frequency smoothing respectively.

\begin{equation}
S_{\text{smooth}}(k, r) = \frac{1}{T} \sum_{t=-T/2}^{T/2} S(k + t, r)
\label{eq:6}
\end{equation}
where T is the window size over which you average the neighboring time frames.
\begin{equation}
S_{\text{smooth}}(k, r) = \frac{1}{F} \sum_{f=-F/2}^{F/2} S(k, r + f)
\label{eq:7}
\end{equation}

with F indicating the the window size over which you average the neighboring frequency bins.

\subsubsection{Cochleagram}

The gammatone filter’s impulse response, employed in creating the cochleagram image representation, using the equation~\ref{eq:8},
\begin{equation}
h(t) = A t^{j-1} e^{-2\pi B t} \cos(2\pi f_c t + \phi)
\label{eq:8}
\end{equation}

In this context, A denotes the amplitude, \(j\) indicates the filter’s order, $B$ defines its bandwidth, and \(f_c\) refers to the center frequency of the filter, $\phi$ corresponding to the phase component, and $t$ represents time~\cite{Patterson1992}. The maximum frequency for the gammatone filters (in Hz) is set to be 20, and the window time, hop time, and the count of channels are set to 0.025, 0.010, and 32, respectively.

The raw audio is downsampled to 16khz before converting the image representation of the audios, the original raw audio’s sampled rate was 22050 hz but in our experimentation, the representations of downsampled audios performed better than the time–frequency representation of the original raw audio of sample rate of 22050 hz.  Since sound produced by majority of the bees stays below 1000 Hz~\cite{Oppenheim1999,abdollahi2022acoustics}, so downsampling the audio didn’t cost us any loss in information and additionally it reduced the memory dependency.

\subsection{Model architecture \& Benchmarking time-frequency Image Representation}

\subsubsection{Feature Extraction }
\begin{figure}[htbp]
    \centering
    \fbox{
    \includegraphics[width=.8\linewidth]{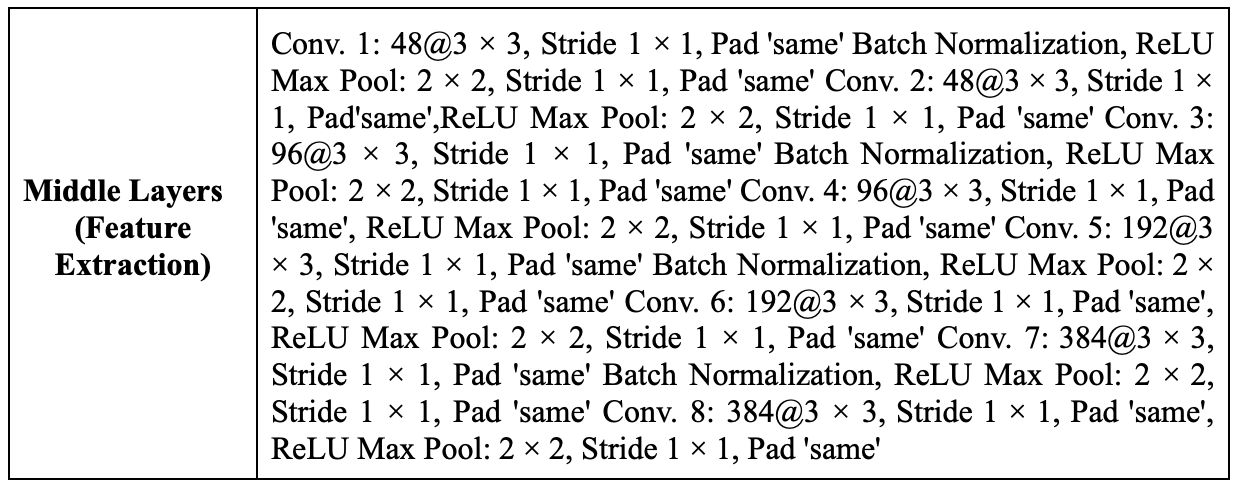}}
    \caption{Architecture of the proposed CNN model used for feature extraction and classification. The network processes time–frequency image representations of beehive audio signals through stacked convolutional and pooling layers, followed by dense layers for multiclass hive state prediction.}
    \label{fig:CNN_FE}
\end{figure}

The CNN architectures and hyperparameter configurations are detailed in  Fig.~\ref{fig:CNN_FE} and Table~\ref{tab:2} respectively. An input image representation with dimensions 64 × 64 × 3 was employed. This network features eight convolutional layers (Conv.\textit{i} where \( \textit{i} \in \mathbb{N} \mid 1 \leq n \leq 8 \)), each followed by ReLU activation, max pooling and batch normalization on every odd number layer as shown in the Fig.~\ref{fig:CNN_FE}.


\begin{table}[htbp]
    \caption{Hyperparameter settings used for training the proposed CNN model on time–frequency image representations of beehive audio signals.}
    \centering
    \begin{tabular}{c c}
        \hline
        \textbf{HyperParameters} &  \textbf{Values} \\
        \hline
        \textbf{Optimization algorithm} &  Rmsprop \\
        \textbf{Max epochs} & 250 \\
        \textbf{learning rate reduction factor} &  0.5 \\
        \textbf{learning rate reduction interval} &   6 \\
        \hline 
    \end{tabular}
    \label{tab:2}
\end{table}


\subsubsection{Classification }
After extraction of the features from the time–frequency representation we fed those features into the shallow artificial neural network (ANN) model. The ANN consists of 2 layers containing 36 and 4 hidden units respectively. The final output  was passed through a softmax function, the softmax function is described in equation~\ref{eq:9},

\begin{equation}
\text{softmax}(z_i) = \frac{e^{z_i}}{\sum_{j=1}^{N} e^{z_j}}
\label{eq:9}
\end{equation}
Where  $z_i$ represents the $i-th$ logit, N is the total number of logits and e is the exponential function. It guarantees that the output probabilities range from 0 to 1 and collectively sum to 1, allowing them to be interpreted as valid probability values for classification tasks~\cite{Bishop2006chapter,Mohamed2012}.

\subsubsection{Benchmarking time-frequency representation }
We trained and evaluated the cnn model with each representation, and the cochleagram representation outperformed all the other representations with respect to accuracy, as presented in Table~\ref{tab:3}.


\begin{table}[htbp]
\centering
\caption{Classification accuracy (\%) of the proposed CNN model using different time–frequency image representations of beehive audio signals.}
\begin{tabular}{ c c }
\hline
\textbf{Representation}    & \textbf{Testing Accuracy} \\ \hline
\textbf{Spectrogram}       & 31.61\%                   \\ 
\textbf{Smoothed-Spectrogram} & 24.11\%                \\ 
\textbf{Mel-Spectrogram}   & 66.66\%                   \\ 
\textbf{Cochleagram}       & \textbf{74.97}\%          \\ \hline
\end{tabular}
\label{tab:3}
\end{table} 

\begin{figure}[h]
    \centering
    \begin{subfigure}[t]{0.48\textwidth}
        \centering
        \fbox{\includegraphics[width=0.95\linewidth, height=3cm]{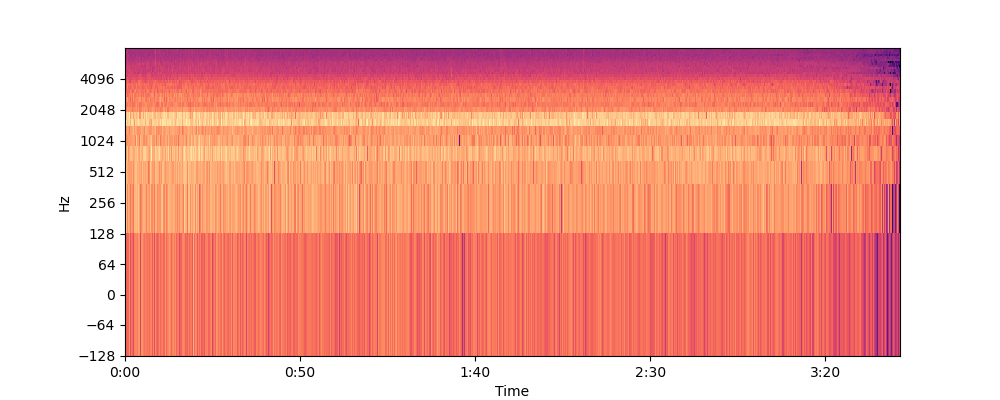}}
        \caption{Queen Not Present}
        \label{fig:queen_not_present}
    \end{subfigure} \hfill
    \begin{subfigure}[t]{0.48\textwidth}
        \centering
        \fbox{\includegraphics[width=0.95\linewidth, height=3cm]{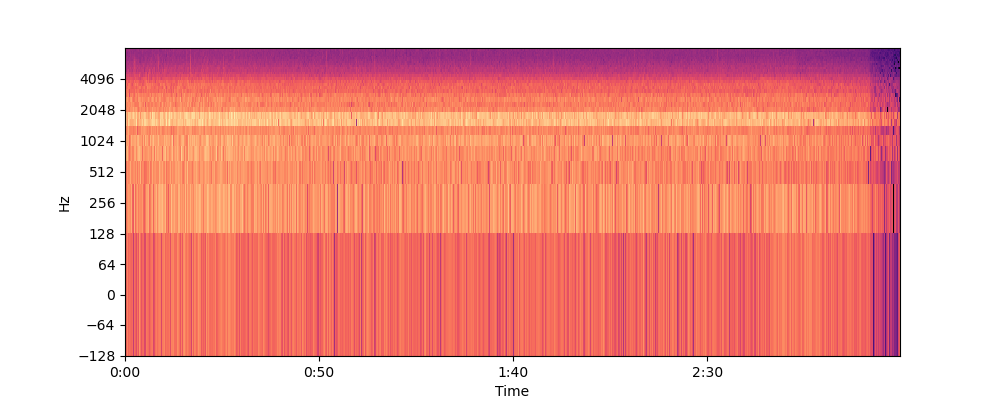}}
        \caption{Queen Present and Newly Accepted}
        \label{fig:present_and_newly_accepted}
    \end{subfigure}

    \vspace{0.5em} 

    \begin{subfigure}[t]{0.48\textwidth}
        \centering
        \fbox{\includegraphics[width=0.95\linewidth, height=3cm]{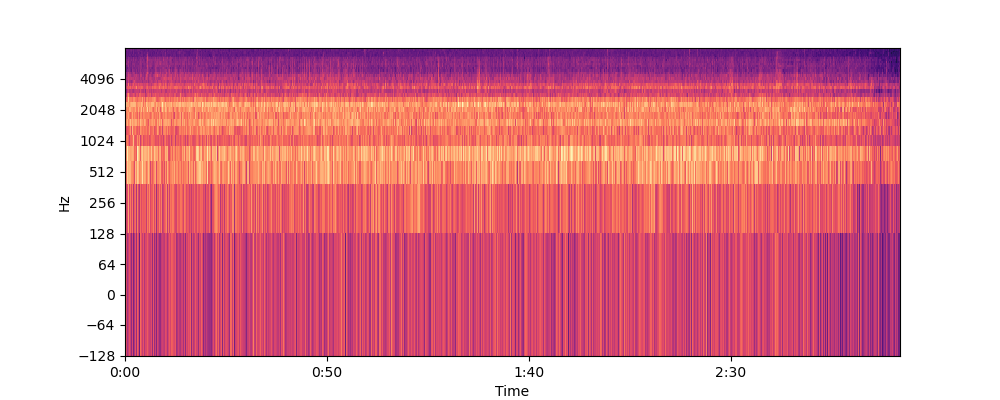}}
        \caption{Queen Present and Rejected}
        \label{fig:present_and_rejected}
    \end{subfigure} \hfill
    \begin{subfigure}[t]{0.48\textwidth}
        \centering
        \fbox{\includegraphics[width=0.95\linewidth, height=3cm]{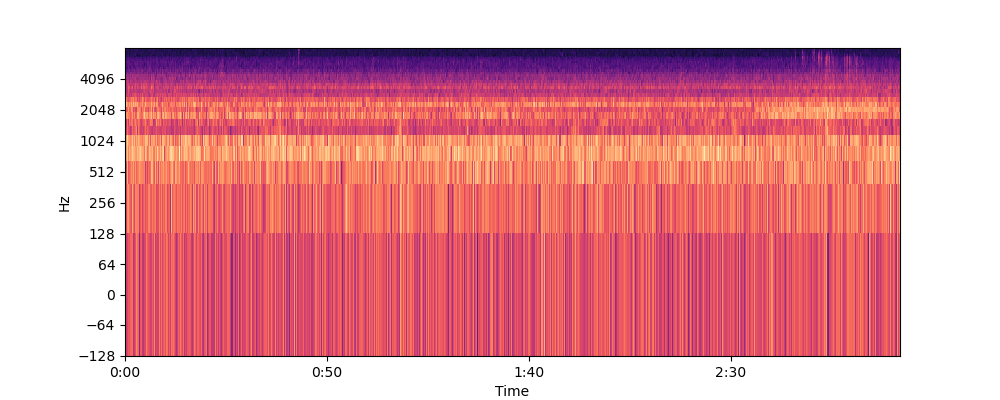}}
        \caption{Queen Present or Original Queen}
        \label{fig:present_or_og}
    \end{subfigure}

    \vspace{0.5em} 

    \caption{\textbf{Cochleagram-based time–frequency image representations:}
    (\textbf{\subref{fig:queen_not_present}}) Queen Not Present,
    (\textbf{\subref{fig:present_and_newly_accepted}}) Queen Present and Newly Accepted,
    (\textbf{\subref{fig:present_and_rejected}}) Queen Present and Rejected,
    (\textbf{\subref{fig:present_or_og}}) Queen Present or Original Queen. These visualizations highlight distinct acoustic patterns used for hive state classification.}
    \label{fig:coch_time_series_rep}
\end{figure}

The cochlea–CNN model performed best among the other time–frequency representations and achieved an accuracy of 74.97\%. After that network tuning, modification and data augmentation by transforming the raw audio files ( e.g. pitch shift, time stretch, and change speed)~\cite{paiva2022} the cochlea–CNN model attained a notable accuracy of 98.31\% on previously unseen test data. The Cochleagram Image Representation of all four beehive states is shown in "Fig~\ref{fig:coch_time_series_rep}``.

\subsection{Training and Validation}
\begin{figure}[htbp]
    \centering
    \includegraphics[width=.5\linewidth]{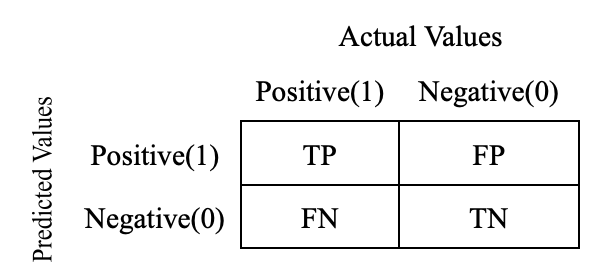}
    \caption{Confusion Matrix Diagram}
    \label{fig:CM_VAL}
\end{figure}

\begin{figure}[h]
    \centering
    \begin{subfigure}[t]{0.48\textwidth}
        \centering
        \fbox{\includegraphics[width=0.95\linewidth, height=5cm]{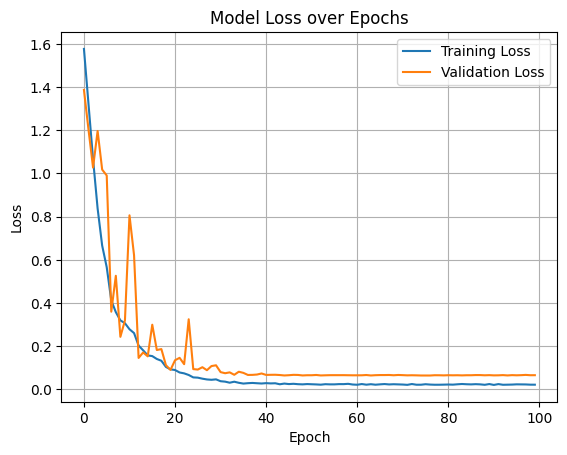}}
        \caption{Loss curve}
        \label{fig:loss}
    \end{subfigure} \hfill
    \begin{subfigure}[t]{0.48\textwidth}
        \centering
        \fbox{\includegraphics[width=0.95\linewidth, height=5cm]{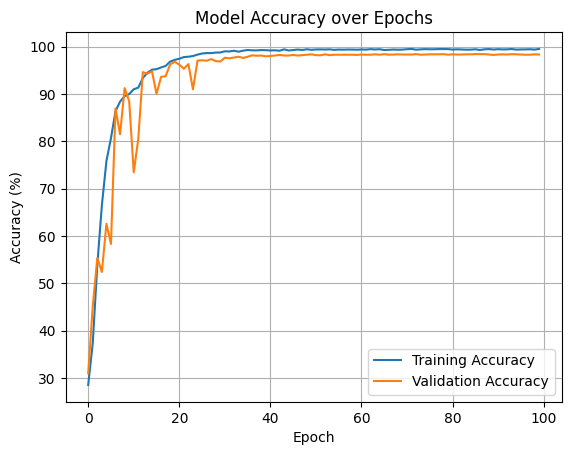}}
        \caption{Accuracy curve}
        \label{fig:acc}
    \end{subfigure}

    \vspace{0.5em}

    \caption{\textbf{Training trends of the CNN model using cochleagram inputs:}
    (\textbf{\subref{fig:loss}}) Training and validation loss,
    (\textbf{\subref{fig:acc}}) Training and validation accuracy.}
    \label{fig:trainig curve}
\end{figure}

We selected the accuracy, F1–score, along with the confusion matrix to assess the classifier’s performance. The following techniques are described in equation~\ref{eq:10},~\ref{eq:11} and Fig.~\ref{fig:CM_VAL} respectively.
\begin{equation}
\text{Accuracy} = \frac{TP + TN}{TP + TN + FP + FN}
\label{eq:10}
\end{equation}

\begin{equation}
F1 = \frac{TP}{TP + 0.5(FP + FN)}
\label{eq:11}
\end{equation}

where \textit{TP, FP, TN} and \textit{FN}   represent true positives, false positives, true negatives, and false negatives, respectively. These methods offer an in-depth assessment of a classification model. Accuracy offers an overall indication of how well the model performs, where F1–score focuses on the minority classes and the confusion matrix that breaks down model outputs into true positives (TP), false positives (FP), true negatives (TN), and false negatives (FN)~\cite{Sokolova2009,Powers2020}. We split the data set with an 85:15 split between training and validation sets, respectively and the count of data samples was 2368.  Fig.~\ref{fig:loss} and ~\ref{fig:acc} illustrate the training and validation loss and accuracy across epochs, respectively.


\subsection{Neural Network Optimization}

For size reduction and inference acceleration we implemented pruning, knowledge distillation and quantization subsequently. And reduced the size of the model by 91.8\% and inference by 66.4\%, while preserving the performance of the model. 

\subsubsection{Pruning}
During the pruning process we performed two types of pruning: random neuron and layer pruning. Random Neuron pruning is the process of removing the neurons from the layer of the network~\cite{Han2015}. And through analysis we’ve seen that those conv layers without batch normalization didn't participate in the learning and classification task. So removing those layers didn't cost any performance and significantly reduced the size of the model~\cite{Liu2017,He2017}. 

\subsubsection{Knowledge Distillation}\label{knowledge distillation}
Knowledge Distillation is also called model distillation, in this process the knowledge is transferred from the bigger model to a smaller and less complex model. This idea was initially introduced by Geoffrey Hinton, et al.~\cite{Hinton2015}, and the objective is to guide the student model to emulate the performance of the teacher model, focusing on achieving computational efficiency and potentially reducing model size. During training, the student model learns from both, the actual labels and the softened outputs generated by the teacher model. The distillation loss encourages the student model to reduce the difference between its predictions and the actual truth labels, while also aiming to match the distribution of predictions generated by the teacher model~\cite{Hinton2015}, as illustrated in ``Algorithm~\ref{alg:knowledge_distillation}''.

Both models use a combination of convolutional layers followed by ReLU activations, batch normalization, and max pooling operations in the feature extraction part, followed by dropout, flatten, and dense layers combined with batch normalization in the classification part. The architecture of both the  models are shown in ``Fig.~\ref{fig:KD}'', the Hidden Layers may not be reduced in the after knowledge–distillation but the number of units/neurons is decreased which decreased the total count of parameters by 40.5\% and the model size reduced from 4.29 MB to 2.56 MB which is nearly 10 times smaller than the earliest model with the almost same accuracy. 

\subsubsection{Quantization}
In Quantization we reduced the precision of a model's weights and activations by converting them from floating-point precision to lower-bit integer formats, such as 16-bit or 8-bit integers~\cite{Nagel2021}. This process significantly reduced the overall model size while accelerating the inference process by reducing the computational load. We encountered some problems during the quantization. As we had to fuse multiple sequential modules (eg: [Conv2d, BatchNorm, ReLU]) into one~\cite{Hubara2020}, but when we tried to fuse the [Conv2d, BatchNorm] layers the accuracy of the model/network decreased drastically. So we tried to calibrate the model after fusing for several epochs, but still the results were the same. We also trained the fused model but the training loss wasn’t converging and the effectiveness of the model remained the same. Since the size of the Distilled network (Student Model) was already small relative to the initial model, so instead of quantizing the whole model/network we only quantized the Classification layer of the model (Dense layer) as nearly 
\(\frac{1}{3}\)  of the total networks is consist of classification layer. The total count of parameters in the student layer were 651,404 and the number of parameters in the classification layer were 262,604, the classification network architecture as outlined in Table~\ref{tab:classification_head}. The count of parameters in the classification layer were calculated using the procedure written below. 


\begin{table}[htbp]
\centering
\caption{Layer-wise configuration of the fully connected classification head of the CNN model. The architecture transforms 4D feature maps into class scores through dropout regularization and two sequential linear layers with batch normalization.}
\label{tab:classification_head}
\begin{tabular}{ll}
\hline
\textbf{Layer} & \textbf{Description} \\
\hline
Dropout 1 & Dropout with probability 0.5 \\
Flatten & Converts 4D tensor to 2D \\
Linear + BatchNorm 1 & Input: $256 \times 4 \times 4$ → Output: 64 units \\
Dropout 2 & Dropout with probability 0.5 \\
Linear + BatchNorm 2 & Input: 64 → Output: 4 (number of classes) \\
\hline
\end{tabular}
\end{table}

\begin{figure}[htbp]
    \centering
    \fbox{\includegraphics[width=1\linewidth]{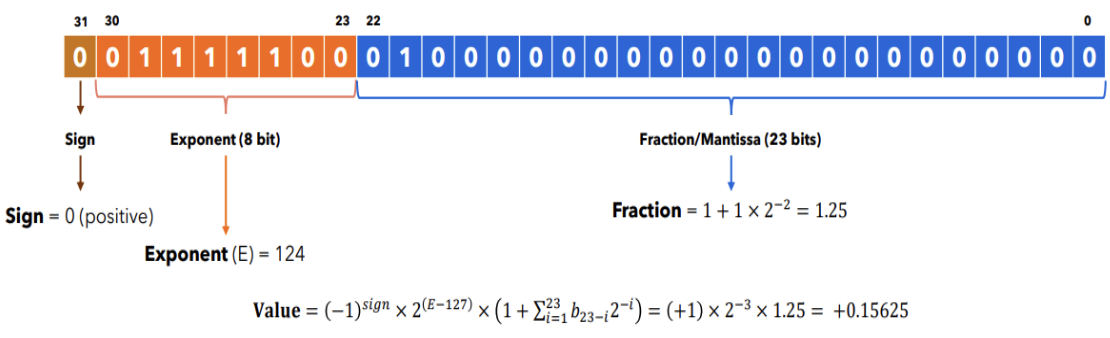}}
    \caption{IEEE 754 single-precision floating-point (Float32) representation format. It consists of a 1-bit sign, 8-bit exponent, and 23-bit mantissa (fraction). This standard format serves as the baseline for model weights and activations before applying lower-bit quantization techniques in neural network optimization~\cite{bib14}.}
    \label{fig:float32}
\end{figure}


The total count of parameters in the given block is 262,604. Since all the parameters of the Distilled network were in float32, the size in megabytes of the 262,604 parameters $\approx$ 1.002 MB floating point numbers representation is shown in Fig.~\ref{fig:float32}. And after quantization, the parameters of the classifier layer are converted to int8 which reduces the size of the classifier layer 4 times.

So after quantizing the distilled network’s classification layer the size compressed to 1.79 mb. This can be calculated by subtracting the storage requirement of the classifier layer from that of the student model, and then adding 25\% of the classifier layer’s size.
\\ 
Size of the classification layer (\(S_{CL}\)) $\approx$ 1.002 MB\\
Size of the Student Model (\(S_{SM}\)) = 2.5 MB\\
Size of the Quantized model = \(S_{SM}\) –\(S_{SL}\)+ {25100 x \(S_{SL}\)}\\
Size of the Quantized model = 2.5 – 1.002 + 0.2505 = 1.75 MB \\

1.75 mb is the calculated storage requirement of the model and 1.79 mb is the observed size of the model which is very close. 
\\
We followed a post-training quantization procedure since we were not quantizing the whole model. First, we determined the range of weights and activations in the classification layer using a calibration dataset~\cite{Jacob2018}. Then, we computed the scale factor S and zero--offset  Z using the equation~\ref{eq:12} \&~\ref{eq:13}, 
\begin{equation}
S = \frac{x_{\text{max}} - x_{\text{min}}}{q_{\text{max}} - q_{\text{min}}}
\label{eq:12}
\end{equation}

\begin{equation}
Z = \text{round}\left(q_{\text{min}} - \frac{x_{\text{min}}}{S}\right)
\label{eq:13}
\end{equation}

where \(x_{\text{min}}\)  and \(x_{\text{min}}\) is the minimum and the maximum values of the original weights/activations. Using these values, each floating-point weight/ activation \(x\) was quantized as in equation~\ref{eq:14},

\begin{equation}
q = \text{round}\left(\frac{S}{x} + Z\right)
\label{eq:14}
\end{equation}

During inference, the quantized weights and activations were dequantized back into floating-point approximations using the equation~\ref{eq:15},

\begin{equation}
x_{\text{approx}} = S \cdot (q - Z)
\label{eq:15}
\end{equation}

This allowed the classification layer to perform computations with reduced precision, minimizing the model’s memory dependencies and computational demands, while retaining the majority of the network’s original performance~\cite{Jacob2018,Krishnamoorthi2018}. By quantizing only the classification layer, we preserved the model's overall accuracy while achieving a significant reduction in inference time and resource consumption~\cite{Krishnamoorthi2018,Wu2020}.

\section{Results}\label{sec:results}
This section outlines the conclusions drawn from the research work and the detailed results of the proposed models are shown and analyzed. The results are organized in way to show the comparative analysis of different time-series representation, their effectiveness with CNN and the impact of models optimization techniques. Each subsection highlights the keys insights, supported by the evaluation metrics employing evaluation measures like accuracy, F1-score, and confusion matrix, to understand the comprehensive understanding of the proposed model performance. Model performance and optimization results and detail analysis is shown section~\ref{result:MP} and~\ref{result:MO} respectively.

\subsection{Model Performance}\label{result:MP}
\begin{figure}[htbp]
    \centering
    \fbox{\includegraphics[width=.7\linewidth]{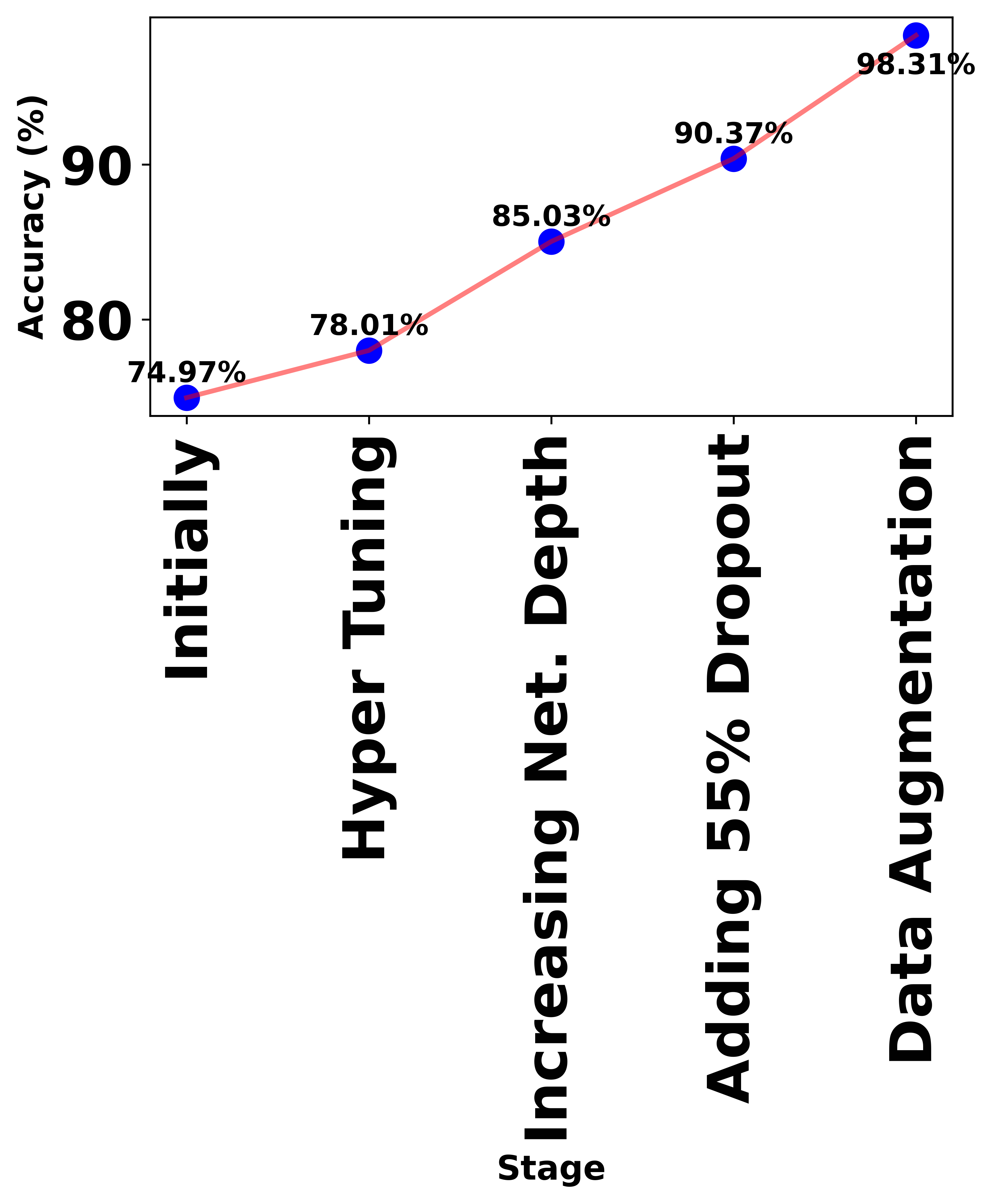}}
\caption{Accuracy progression across key model development stages for the CNN using cochleagram inputs. Each stage—initial setup, hyperparameter tuning, increased network depth, dropout regularization, and data augmentation—demonstrates incremental improvements, culminating in a peak accuracy of 98.31\%.}
    \label{fig:3}
\end{figure}

Initially the use of cochleagram image representation resulted in superior classification performance in comparison to the other time-frequency representations achieving as accuracy of 74.97\% as shown in Table~\ref{tab:3}. And after model hyper parameter tuning the accuracy of the model increase by 4.05\% and achieved the accuracy of 78.01\%. Then we increased the depth of the neural network by increasing the count of convolution layer followed by batchnorm layer and max pooling layer which improved the model's overall accuracy by ~9\% and achieved the accuracy of 85.03\%. After that we regularize the model by adding 55\% dropout in the classification layer which improve the model's accuracy by 6.28\% and achieved the accuracy of 90.37\%. And at the end we performed data augmentation on the raw audio data which substantially enhanced the model’s performance by 8.8\% and pushed the model performance to 98.31\% test accuracy. The progress graph is shown in ``Fig.~\ref{fig:3}'', illustrates the progressive improvement in accuracy through various stages of model refinement. Each stage significantly boosted the accuracy from an initial 74.97\% to a remarkable 98.31\% after implementing data augmentation.

The classification report of the best performing model can be observed in Table~\ref{tab:5}, which indicates the balanced performance of the model across all classes that are Queen Not Present, Queen Present and Newly Accepted, Queen Present and Rejected, Queen Present or Original Queen. The class "Queen not present" achieved a precision value of 0.98, a recall value of 0.99 along with an F1-score of 0.99. The "Queen present and newly accepted" class recorded a marginally reduced recall recall of 0.97 but maintained high precision (0.99) and F1-score (0.98). The "Queen present and rejected" class yielded consistent scores of 0.97 across all three metrics. Notably, the "Queen present or original queen" class achieved nearly perfect classification with a recall of 1.00 and F1-score of 0.99.

\begin{figure}[htbp]
    \centering
    \fbox{\includegraphics[width=.9\linewidth]{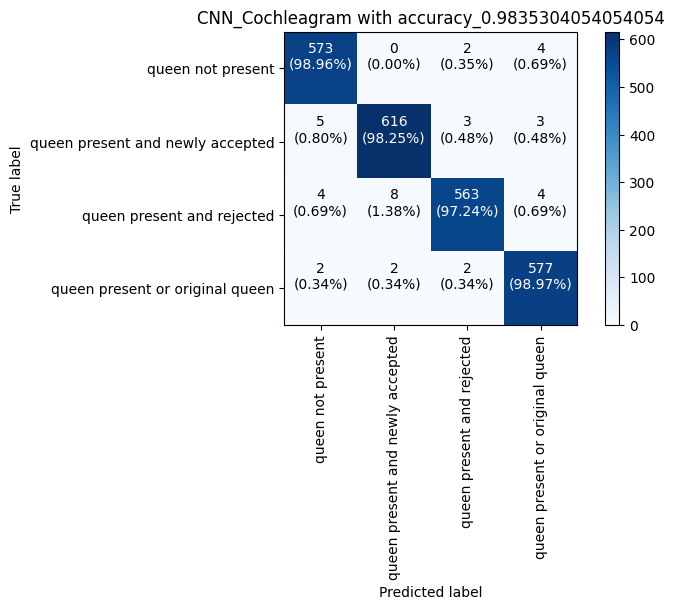}}
    \caption{Confusion matrix of the best-performing CNN model trained on cochleagram representations for beehive state classification. The model achieves an overall accuracy of 98.35\%, with high precision and minimal misclassification across all four hive state categories}
    \label{fig: 4}
\end{figure}

To further assess the model performance and class-wise behavior, we analyzed the confusion matrix is depicted in ``Fig.~\ref{fig: 4}'' on unseen data. The model demonstrated particularly high accuracy in distinguishing between the classes, correctly identifying 98.96\% of the “queen not present” samples, 98.25\% of “queen present and newly accepted,” 97.24\% of “queen present and rejected,” and 98.97\% of the “queen present or original queen” samples. Most of the misclassifications were limited to acoustically similar categories, such as confusion between the “newly accepted” and “rejected” classes. This shows the model's capacity to generalize effectively, while capturing the subtle differences in beehive acoustic patterns.

\begin{table}[htbp]
\centering
\caption{Class-wise precision, recall, and F1-score of the final CNN model trained on cochleagram representations for multiclass beehive state classification. The model demonstrates consistently high performance across all four hive states.}
\begin{tabular}{l c c c c }
\hline
\textbf{Class}                               & \textbf{Precision} & \textbf{Recall} & \textbf{F1-Score} & \textbf{Support} \\
\hline
Queen not present                            & 0.98               & 0.99            & 0.99              & 549              \\ 
$Q_P$ and newly accepted             & 0.99               & 0.97            & 0.98              & 627              \\ 
$Q_P$ and rejected                   & 0.97               & 0.97            & 0.97              & 583              \\ 
$Q_P$ or original queen              & 0.99               & 1.00            & 0.99              & 609              \\
\textbf{Accuracy}                            &                    &                 & \textbf{0.98}     & 2368             \\ 
\textbf{Macro avg}                           & 0.98               & 0.98            & 0.98              & 2368             \\ 
\textbf{Weighted avg}                        & 0.98               & 0.98            & 0.98              & 2368             \\ 
\hline
\label{tab:5}
\end{tabular}
\end{table}

\subsection{Model Optimization}\label{result:MO}
Subsequent to enhancing its performance, the model required 21.82 MB of storage space. Then we simplified the network by optimizing it using optimization techniques such as pruning, knowledge distillation and Quantization. At first we converted the RGB cochleagram time-series representation to grayscale representation and change the cochleagram-cnn input from 3 dimensional input to single dimensional input, which led to a marginal size reduction but yielded a 0.1\% improvement in model accuracy after calibration, indicating that grayscale input had minimal impact on performance while maintaining similar size and inference time. 

Subsequent pruning including neuron and layer pruning reduce the size of the model 48.73\% and 61.63\% respectively and reduce the overall size from 21.81 mb to 4.29 mb which is 80.33\% reduction in the size and 41.67\% reduction in inference time from 5.88 sec to 3.43 sec, while maintaining a reasonable accuracy of 97.00\%. 

Further gains were observed through knowledge distillation and quantization, where model was compressed to just 2.50 MB with a modest accuracy of 97.09\% and a significantly faster inference time of 2.13 seconds. And quantization reduced the size to 1.79 MB and inference time to 1.91 seconds,exhibiting a mere 1.3\% reduction in accuracy when benchmarked against the original

We successfully compressed the model 91.8\% reduced it's size from 21.82 mb to 1.79 mb and accelerated the inference by 66.4\% without losing much of the overall competency of the model. The Table~\ref{tab:6} summarizes the model’s overall efficacy on multiple stages of optimization considering accuracy, size, inference time and number
of parameters.

\begin{table}[htbp]
\centering
\caption{Comparison of the baseline CNN model and its optimized variants in terms of classification accuracy, model size, inference time, and number of parameters. Optimizations include grayscale preprocessing, pruning, and distillation. For the quantized model, only the final classification layer was quantized to reduce memory usage while maintaining performance.}
\begin{tabular}{l c c c c}
\hline
\textbf{Model}             & \textbf{Accuracy (\%)} & \textbf{Size (MB)} & \textbf{Time (s)} & \textbf{Parameters} \\
\hline
CNN\_Tensorflow            & 98.31                  & 21.82              & 5.69              & 5,719,690           \\
Grayscale\_Tensorflow      & 98.41                  & 21.81              & 5.88              & 5,717,962           \\
Neuron Pruned              & 96.75                  & 11.18              & 7.20              & 2,858,224           \\
Layer Pruned               & 97.00                  & 4.29               & 3.43              & 1,094,952           \\
Distilled Model            & 97.09                  & 2.50               & 2.13              & 651,404             \\
Quantized Model            & 97.00                  & 1.79               & 1.91              & 651,404             \\
\hline
\label{tab:6}
\end{tabular}
\end{table}



\begin{figure}[!h]
    \centering
    \begin{subfigure}[t]{0.48\textwidth}
        \centering
        \fbox{\includegraphics[width=0.95\linewidth, height=5cm]{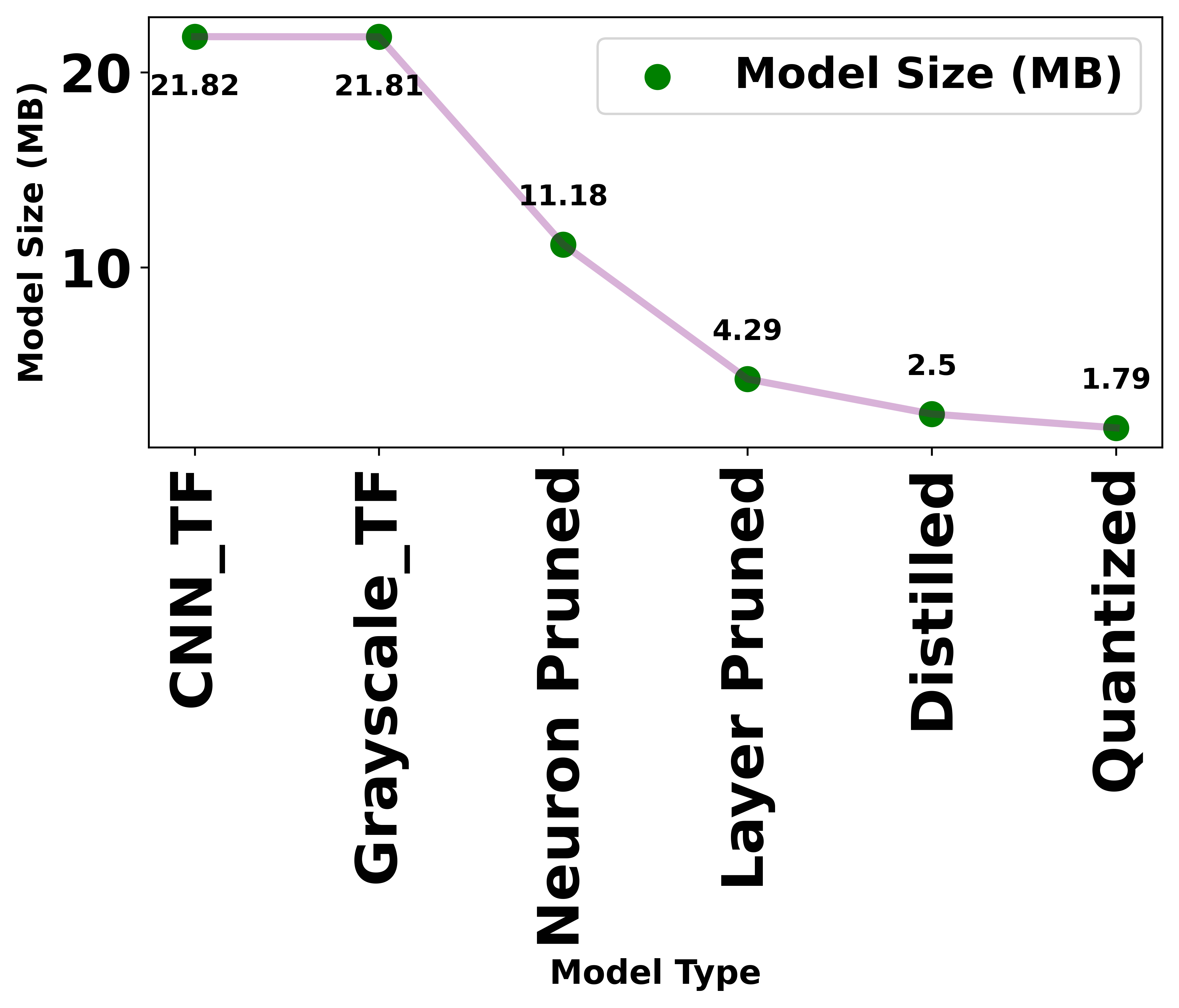}}
        \caption{} \label{fig:5a}
    \end{subfigure}
    \hfill
    \begin{subfigure}[t]{0.48\textwidth}
        \centering
        \fbox{\includegraphics[width=0.95\linewidth, height=5cm]{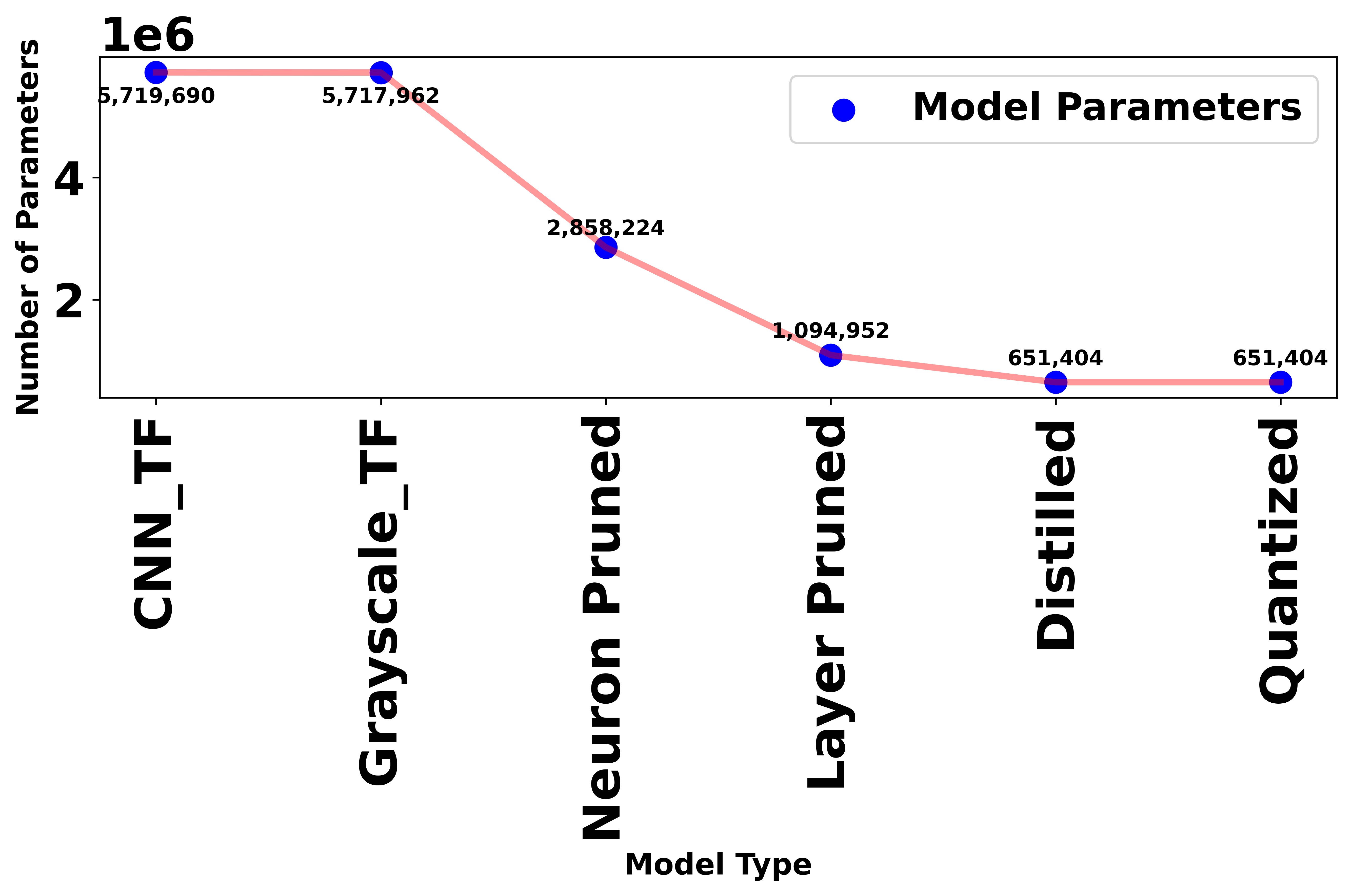}}
        \caption{} \label{fig:5b}
    \end{subfigure}
    \hfill
    \begin{subfigure}[t]{0.48\textwidth}
        \centering
        \fbox{\includegraphics[width=0.95\linewidth, height=5cm]{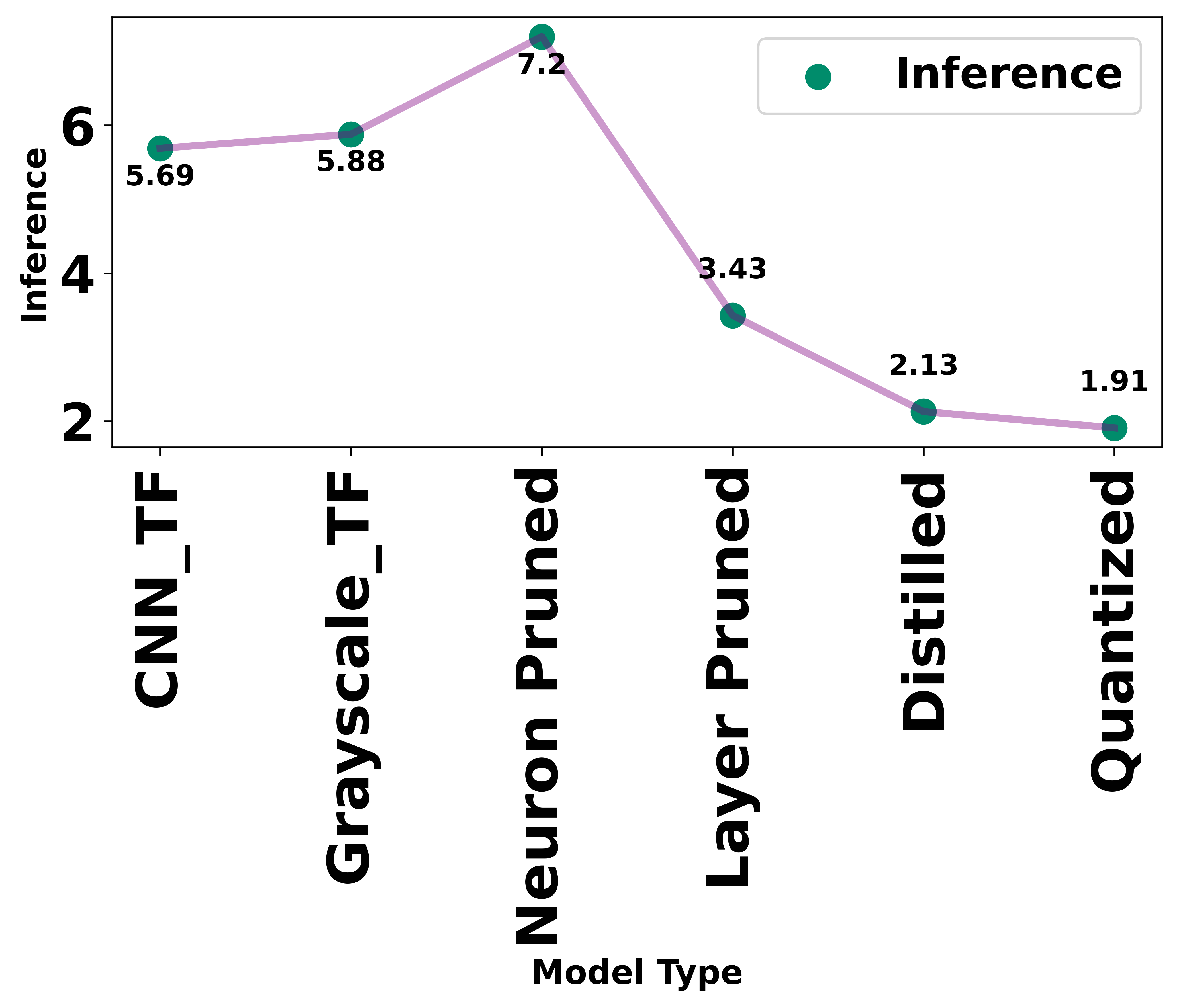}}
                \caption{} \label{fig:5c}
    \end{subfigure}
    \hfill
    \begin{subfigure}[t]{0.48\textwidth}
        \centering
        \fbox{\includegraphics[width=0.95\linewidth, height=5cm]{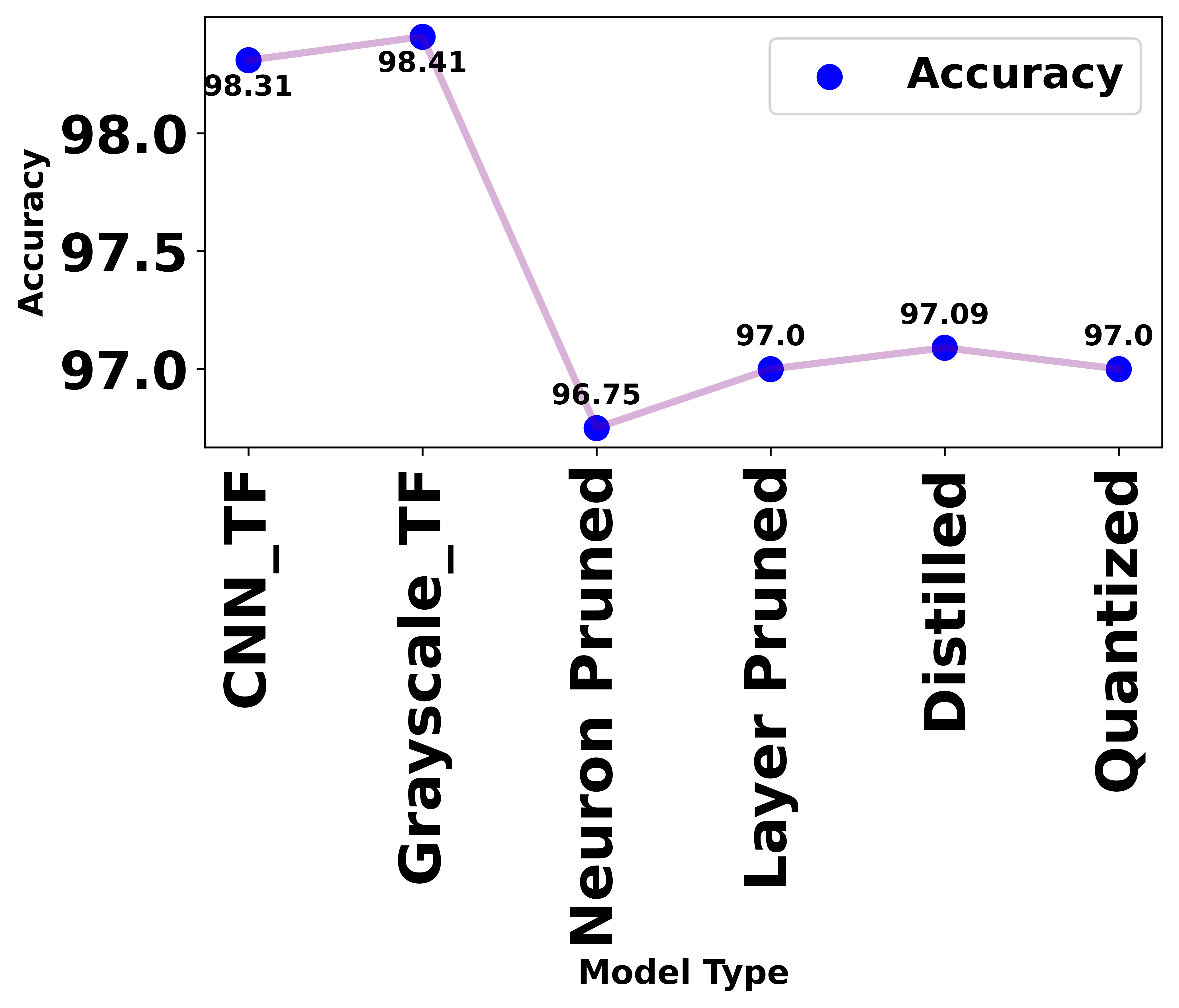}}
        \caption{} \label{fig:5d}
    \end{subfigure}

    \caption{
    (\subref{fig:5a})Optimization Stages vs Network's Size (in MB)
    (\subref{fig:5b})Optimization Stages vs Number of Parameters 
    (\subref{fig:5c}) Optimization Stages vs inference time (in seconds)
    (\subref{fig:5d}) Optimization Stages vs Accuracy
}
    \label{fig:shap_and_cfm}
\end{figure}

Figure~\ref{fig:shap_and_cfm} shows how different optimization stages affect model size, parameter count, inference time, and classification accuracy. In Figure~\ref{fig:5a}, the baseline CNN\_TF model is 21.82~MB. This size stays the same after grayscale preprocessing but drops significantly to 11.18~MB after neuron pruning and to 4.29~MB after layer pruning. Knowledge distillation then reduces the size to 2.50~MB, and partial quantization compresses it further to 1.79~MB. This represents a total size reduction of 91.8\% compared to the baseline.

Figure~\ref{fig:5b} shows a similar trend in parameter count. The baseline starts with 5.72~million parameters, which decreases to 2.85~million after neuron pruning and to 1.49~million after layer pruning. Distillation creates a smaller architecture with 651{,}000 parameters, but there is no further reduction from quantization since only the classification layer weights are quantized.

In terms of inference time (Figure~\ref{fig:5c}), the baseline model takes 5.69~seconds. This increases slightly to 5.88~seconds after grayscale conversion and peaks at 7.20~seconds after neuron pruning due to irregular sparsity overhead. Layer pruning improves this by cutting inference time to 3.43~seconds. Distillation and quantization further reduce the inference time to 2.13~seconds and 1.91~seconds, respectively, achieving a 66.4\% reduction in inference time compared to the baseline model.

Lastly, Figure~\ref{fig:5d} shows the accuracy trade-offs. The baseline accuracy is 98.31\%, which slightly improves to 98.41\% after grayscale conversion but falls to 96.75\% after neuron pruning. Layer pruning brings performance back to 97.0\%, with distillation achieving 97.09\% and quantization maintaining 97.0\%. This shows that significant compression and runtime improvements can be reached with only a 1.3\% absolute accuracy loss compared to the uncompressed baseline model.

\begin{figure}[!h]
    \centering
    \fbox{\includegraphics[width=.75\linewidth]{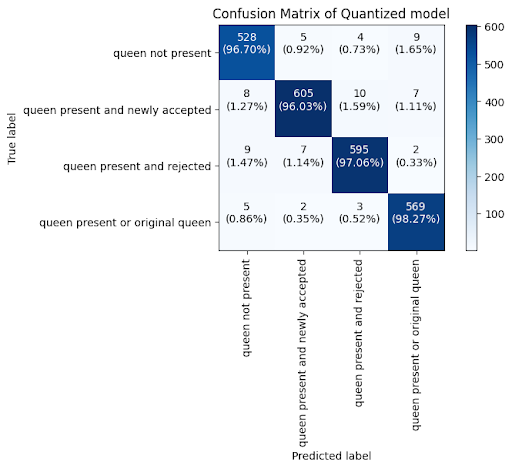}}
    
\caption{Confusion matrix of the final optimized CNN model for beehive state classification. The model was compressed using pruning, knowledge distillation, and quantization of the classification layer. It maintains high predictive accuracy across all four classes with minimal misclassification.}
    \label{fig:optimized_cm}
\end{figure}

The confusion matrix of the final optimized model is shown in ``Fig.~\ref{fig:optimized_cm}'', revealing the model’s stable and balanced classification results across all categories.

These results highlight the superior trade-off achieved through optimizing using techniques like pruning, knowledge distillation and quantization, offering substantial reductions in memory and latency without significant loss in classification performance.


\section{Discussion}\label{sec:discussion}

In this research the effectiveness of CNNs in classifying complex bioacoustic signals to monitor beehive states has been shown, and also the importance of time-series representation of an audio signal. By evaluating the performance of four varied time–frequency image representation methods, including Spectrogram and Mel-Spectrogram, Smoothed-Spectrogram, and Cochleagram with CNN, we identified that the Cochleagram as the most informative and suitable representation for CNN-based classification scenarios. The final model produced an impressive accuracy of 98.31\% on unseen test dataset, after subsequent refinements including  network architectural enhancements and data augmentation on raw audio data.

Our findings align with the previous researches, that suggest the importance of time-series image representation of for beehive classification using CNN. However none of them explored or demonstrated the superior performance of the Cochleagram time-series representation in beehive classification task. Previous studies were largely focused on binary classification (e.g., queen present vs. not present) and relied on representations such as MFCCs or Mel-Spectrograms. 

In contrast, our research shows the feasibility and advantages of cochleagram time-series image representation which improved the CNN performance in multiclass-classification. These four class (Queen Not
Present, Queen Present and Newly Accepted, Queen Present and Rejected, Queen
Present or Original Queen) are both practical and beneficial in beehive monitoring. This more detailed classification provides deeper insights of the hive and has the potential to significantly improve real-time beehive monitoring.

This paper also addresses one of the critical barrier to real-time beehive monitoring i.e model size and inference time. Although Deep Convolutional Neural Network models often shows high classification accuracy, their practical application in field settings is limited due to computational and memory constraints which also makes the hardware expensive. Though the optimization techniques such as pruning, knowledge distillation and quantization, we reduce the CNN model size by 91.8\% and reduce the inference time by 66.4\% with only marginal degradation in performance (final accuracy of 97.00\%), which reduce the dependency of high end hardware and also reduces the cost of the overall system. This highlights a favorable trade-off between computational efficiency and predictive power.

Notably, the evaluation metrics such as confusion matrices and classification report which provides the class-wise evaluation, shows that the model's balanced performance across all four beehive states, with majority of misclassifications occurring between acoustically similar states. This indicates the model’s capability to capture nuanced acoustic variations in hive behavior, an essential aspect for accurate hive health monitoring.

An identified shortcoming in this study is the utilization of a single publicly available dataset collected from a specific geographic region. This may impact the generalizability of the model to hives under different environmental conditions or bee species. Future work can address this by incorporating diverse datasets and exploring domain adaptation techniques. Moreover, real-time deployment and field testing are needed to evaluate robustness under operational constraints such as background noise, hardware limitations, and temporal drifts in data.

In summary, this research not only advances the state of audio-based beehive monitoring but also sets a practical foundation for scalable, real-time applications through lightweight CNN architectures.

\section{Conclusion}\label{sec:conclusion}

This research focuses on a holistic method of classifying beehives states based on Convolutional Neural Networks (CNNs) applied to bioacoustic signals. Through a systematic study of four time–frequency image representations, we found that the Cochleagram time-series representation is the most suitable input format to facilitate CNN performance using bioacoustic data. This representation enabled a successful classification model for beehive multi state classification, while in previous research only the binary state classification was applied.

The proposed model outperformed prior approaches  with an accuracy of 98.31\%  on the testing samples with a strong generalization across all the four classes or states. In addition, we addressed the major obstacles encountered during deployment of models and real-time monitoring —size and inference latency—by applying optimization techniques  involving pruning, knowledge distillation, and quantization. With these optimizations, the size and inference time of the model was reduced by 91.8\% and  66.4\% respectively,  while maintaining high performance. The optimized model exhibited only a minor accuracy drop of 1.33\%, achieving 97.00\% accuracy on unseen data, thus allowing the model  to be deployed to real-time and resource-limited scenarios.

This research enhances not only extending past the contemporary state-of-the-art approaches in hive monitoring, but also provides a deployable, scalable and cost-effective solution for both researchers and beekeepers. The methodology  described here can be applied to other environmental or agricultural audio classification tasks.

Future work should verify the model’s efficacy on more diverse datasets and under real-time/real-world conditions aiming to improve the robustness and generalization of the model. Incorporating external factors such as temperature, humidity, and weather data, which could further enhance the classification performance and practical utility.


\section{Acknowledgements}
The author(s) acknowledge and are thankful to Council of Scientific and Industrial Research–Central Scientific Instruments Organisation (CSIR–CSIO), Sector 30, Chandigarh, India for provind technical support for conducting the study under the project GAP00492.

\section*{Declarations}

\begin{itemize}

  \item \textbf{Data availability} \\
    The dataset used in this study is publicly available at: \url{https://www.kaggle.com/datasets/annajyang/beehive-sounds}.

  \item \textbf{Materials availability} \\
   The computational experiments were conducted on a local machine equipped with an Intel Core i9 (12th Gen) processor, 32 GB RAM, and an octa-core CPU. Python 3.9 was used for implementation, along with open-source libraries such as TensorFlow, Keras, scikit-learn, librosa for audio processing, and Matplotlib, Seaborn, Google Sheets for visualizations. No specialized hardware or proprietary software was used.

   \item \textbf{Source code availability} \\
   The source code available at \href{https://github.com/CSIO-FPIL/beehive-classification}{https://github.com/CSIO-FPIL/beehive-classification}.

  \item \textbf{Author contribution} \\
  Harshit: Conceptualization, Methodology, Experimentation, Writing – original draft, Visualization. Rahul Jana: Discussion, Writing – review \& editing. Dr. Ritesh Kumar: Supervision.
\end{itemize}


\bibliography{sn-bibliography}

\clearpage
\begin{appendices}

\section{Detailed Architectures of Teacher and Student Models with Distillation Pseudocode}\label{secA1}

This appendix provides additional visual and algorithmic details related to the knowledge distillation optimization process discussed in Section~\ref{knowledge distillation}.

\begin{figure}[htbp]
    \centering
    \includegraphics[width=01\linewidth]{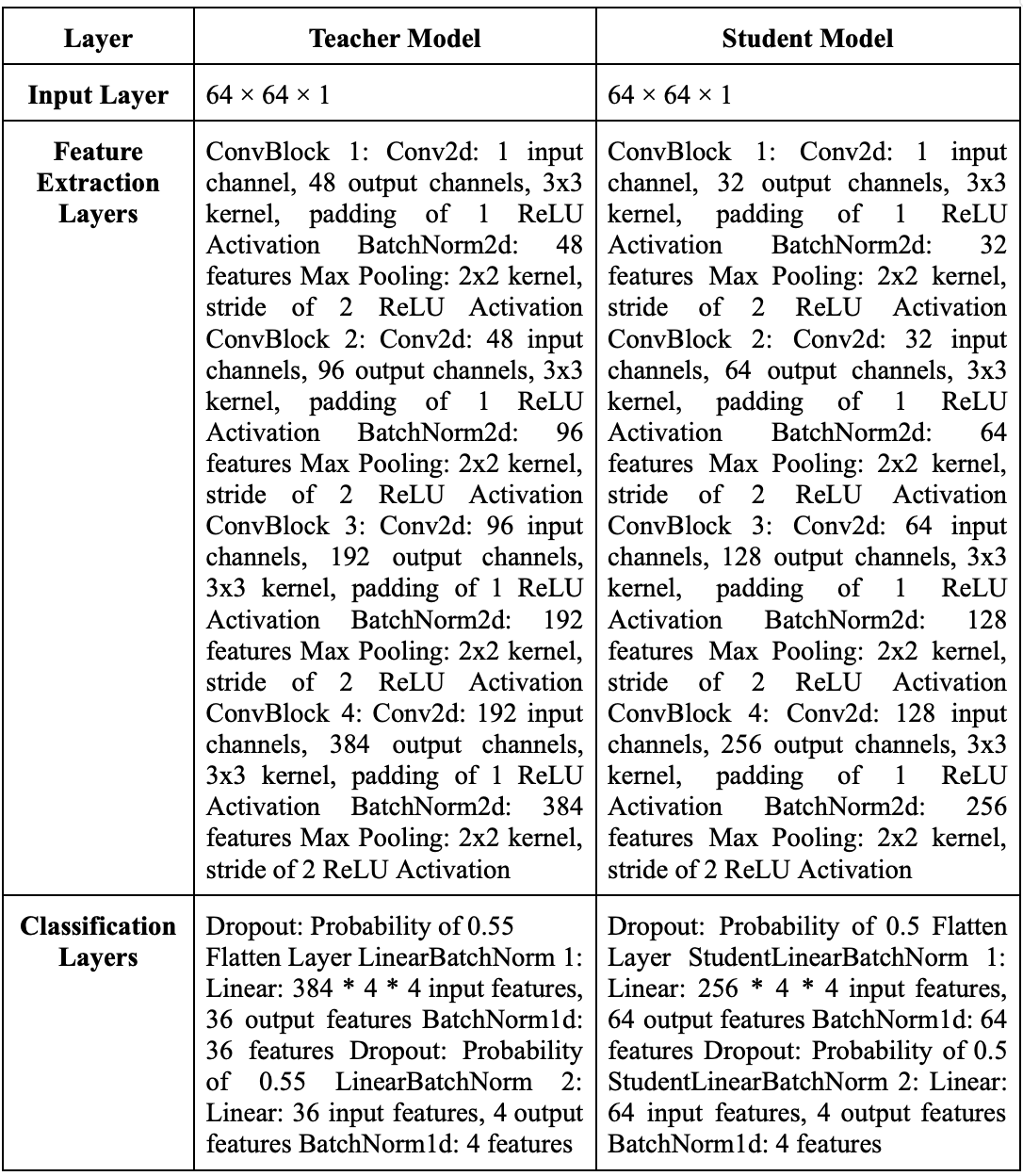}
    \caption{Network Architecture of Teacher and Student Model}
    \label{fig:KD}
\end{figure}


\begin{algorithm}[htbp]
\caption{Knowledge Distillation Training Procedure}
\label{alg:knowledge_distillation}
\textbf{Inputs:} \\
\noindent
\begin{tabular}{>{\raggedright\arraybackslash}p{0.49\linewidth} >{\raggedright\arraybackslash}p{0.45\linewidth}}
$T\_Model$: Teacher model (pre-trained) & 
$S\_Model$: Student model (untrained) \\
$D\_train$: Training dataset  & 
$\eta$: Number of training epochs \\
$\Gamma$: Temperature for softening probabilities & 
$\alpha$: Weight for distillation loss \\
$\beta$: Weight for ground truth loss & 
$optimizer$: Optimization algorithm (e.g., SGD, Adam) \\
$distill\_loss$: Cross-entropy loss between student and teacher soft targets & 
$gt\_loss$: Cross-entropy loss between student predictions and true labels \\
\end{tabular}

\begin{algorithmic}[1]

\Function{ForwardPass}{Model, X}
    \State logits $\gets$ Model(X)
    \State \Return logits
\EndFunction

\Function{SoftmaxWithTemperature}{logits, $\Gamma$}
    \State exp\_logits $\gets \exp(\frac{logits}{\Gamma})$
    \State probs $\gets \frac{exp\_logits}{\sum exp\_logits}$
    \State \Return probs
\EndFunction

\Function{CrossEntropyLoss}{pred\_probs, true\_probs}
    \State loss $\gets - \sum (true\_probs \cdot \log(pred\_probs))$
    \State \Return loss
\EndFunction

\Function{DistillationLoss}{S\_probs, T\_probs}
    \State distill\_loss $\gets$ \Call{CrossEntropyLoss}{S\_probs, T\_probs}
    \State \Return distill\_loss
\EndFunction

\Function{CalculateTotalLoss}{S\_logits, y, T\_probs, $\alpha$, $\beta$, $\Gamma$}
    \State S\_probs\_distill $\gets$ \Call{SoftmaxWithTemperature}{S\_logits, $\Gamma$}
    \State distill\_loss $\gets$ \Call{DistillationLoss}{S\_probs\_distill, T\_probs}
    \State S\_probs\_gt $\gets$ \Call{SoftmaxWithTemperature}{S\_logits, 1}
    \State gt\_loss $\gets$ \Call{CrossEntropyLoss}{S\_probs\_gt, y}
    \State total\_loss $\gets \alpha \cdot$ distill\_loss $+ \beta \cdot$ gt\_loss
    \State \Return total\_loss
\EndFunction

\For{epoch = 1 to $\eta$}
    \For{each batch (X, y) in $D\_train$}
        \State \textbf{with no gradient:}
        \State T\_logits $\gets$ \Call{ForwardPass}{T\_Model, X}
        \State T\_probs $\gets$ \Call{SoftmaxWithTemperature}{T\_logits, $\Gamma$}
        \State S\_logits $\gets$ \Call{ForwardPass}{S\_Model, X}
        \State total\_loss $\gets$ \Call{CalculateTotalLoss}{S\_logits, y, T\_probs, $\alpha$, $\beta$, $\Gamma$}
        \State optimizer.zero\_grad()
        \State total\_loss.backward()
        \State optimizer.step()
    \EndFor
\EndFor

\end{algorithmic}
\end{algorithm}

\end{appendices}

\end{document}